\documentclass[preprint,12pt]{elsarticle}

\usepackage{graphicx}
\usepackage{amssymb}
\usepackage{bm}
\usepackage{amsmath}

\journal{Journal of Computational Physics}

\begin{document}

\begin{frontmatter}



\title{Stable Numerical Integration of an Epitaxial Growth Model with 
Slope Selection}


\author[colby]{Gregory M. Seyfarth}
\author[bucknell]{Benjamin P. Vollmayr-Lee\corref{corr}}
\cortext[corr]{Corresponding author}

\address[colby]{Department of Physics and Astronomy, Colby College, 
   Waterville, Maine 04901, USA}

\address[bucknell]{Department of Physics and Astronomy, Bucknell University, 
  Lewisburg, Pennsylvania 17837, USA}

\begin{abstract}
We consider a continuum phase field model for crystal growth via
molecular beam epitaxy, with the goal of determining stable numerical
time integration methods for the dynamics.  We parametrize a class of
semi-implicit methods that are linear in the updated field, which
allows for efficient implementation with fast Fourier transforms.  We
perform unconditional von Neumann stability analysis to identify the
region of stability in parameter space, and then test these
predictions numerically for gradient stability.  We find strong
agreement between the approaches.
\end{abstract}


\end{frontmatter}


\section{Introduction}
\label{Introduction}

In growing crystal surfaces by molecular beam epitaxy (MBE), the
Ehrlich-Schwoebel-Villain effect
\cite{Ehrlich66,Schwoebel66,Villain91} can destabilize a flat
interface and lead to the formation of pyramids or mounds (see
\cite{Misbah10} for a recent review).  These surface features then
coarsen, with their height and spatial extent growing as powers of
time. Theoretical studies of MBE coarsening typically employ continuum
models, justified by appeal to the large distance and slow time scales
involved.  The resulting field equations of motion are nonlinear, and
to make progress they must be integrated numerically, a process which,
unfortunately, is hampered by numerical instabilities.  As such, much
recent effort has been devoted to finding stable integration methods.
In this work, we derive a class of stable numerical integration
methods that are particularly efficient and simple to implement
because the updated field can be obtained via the fast Fourier
transform (FFT).

The model we consider employs a height field $h(x,y,t)$ that is a
continuous function of space and time, and which obeys the equation of
motion
\begin{equation}
\frac{\partial h}{\partial t} = -\nabla^4 h
 - \bm\nabla \cdot \Bigl\{(1- |\nabla h|^2) \bm \nabla h \Bigr\},
\label{eq:eq_of_motion}
\end{equation}
applicable for homoepitaxial growth with isotropic slope selection.  The
motivation for this and related models is discussed below.  With these
dynamics, equilibrated regions of uniform gradient and unit slope
form.  Domains with different slope orientations meet at edges of
constant width, and as the system evolves the edges are healed out,
resulting in the growth of the characteristic domain size.  For this
particular model it has been found from theoretical analysis
\cite{Ortiz99,Moldovan00,Biagi12}, simulations \cite{Moldovan00,Wang10}, and
rigorous bounds \cite{Kohn03} that the characteristic domain size
$L(t)$ grows with time as $L\sim t^{1/3}$. 

Numerical simulations of coarsening are useful for testing scaling and
the predicted growth laws and for measuring properties of the scaling
state, such as correlations, growth law amplitudes, autocorrelation
functions, and more (see \cite{Bray94} for a coarsening review).  But
these simulations face several restrictions.  To reach the asymptotic
scaling regime, it is necessary to evolve until $L(t) \gg w$, where
$w$ is the width of the edges.  But the lattice size $\Delta x$ must
be sufficiently smaller than the edge width in order to resolve the
edge shape and corresponding line tension.  Finally, the system size
$L_\text{sys}$ must be large enough that domains can grow into the
scaling regime before finite size effects appear.  To satisfy this
string of conditions, $\Delta x \ll w \ll L(t) \ll L_\text{sys}$,
requires lattices of very large linear size $L_\text{sys}/\Delta x$,
evolved to late times.

For this reason, it is desirable to use integration schemes that are
\textit{accuracy}-limited rather than \textit{stability}-limited.
Euler integration of Eq.~(\ref{eq:eq_of_motion}) is only stable for
time steps $\Delta t$ smaller than a threshold determined by the
lattice spacing.  In contrast, an unconditionally stable method,
i.e.,\ one with no conditions on $\Delta t$, would allow a time step
determined by the natural time scale of the dynamics, which turns out
to be considerably more efficient.  Accuracy considerations require
the typical distance traveled by an edge within one step to be held
fixed \cite{Vollmayr-Lee03,Cheng05}, and since the characteristic edge
velocity scales as $v_\text{edge}\sim \partial L/\partial t \sim
t^{-2/3}$, this allows a growing time step $\Delta t\sim t^{2/3}$.
Using $dt/dn\sim \Delta t$, where $n$ is the number of integration
steps, it follows that unconditionally stable methods allow accurate
evolution with $t\sim n^3$, rather than the stability-limited $t\sim
n$.  For typical simulation parameters, this provides greater than a
$1000$-fold increase in efficiency!

Eyre provided a general approach for generating unconditionally stable
semi-implicit integration methods, based on a splitting into expansive
and contractive terms \cite{Eyre98b}.  Wang, Wang, and Wise used this
approach for Eq.~(\ref{eq:eq_of_motion}), as well as an MBE model
without slope selection \cite{Wang10}, and this approach has now been
extended to a second-order in time method \cite{Shen12} and other
developments \cite{Qiao12,Chen12b,Chen13}.  These
schemes are \textit{gradient stable}, which means they preserve the
energy-decreasing property of the continuous-time equation.  However,
these Eyre-based schemes have the drawback that usually a nonlinear term 
must be treated implicitly, requiring an iterative method
to find the updated field, and in the worst case no guarantee of
convergence or a unique solution.  An alternate approach is to
restrict consideration to steps with linear implicit terms that can be
solved directly by FFT, determine the range of step parameters that
satisfy unconditional von Neumann (UvN) stability, and then test
these parameters numerically for gradient stability.  This approach
yielded stable, direct steps for the Cahn-Hilliard and Allen-Cahn
equations \cite{Vollmayr-Lee03}, and it is the program we follow here
for the MBE model.

\begin{figure}
\begin{center}
\includegraphics[width=3.25in]{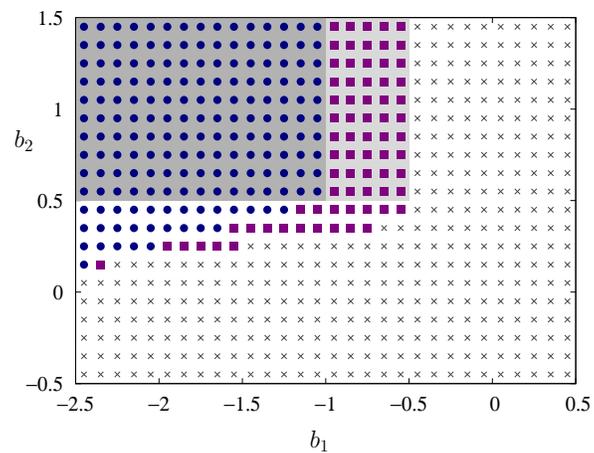}
\caption{(color online) Stability diagram for the parameters $b_1$ and
  $b_2$ in Eq.~(\ref{eq:discrete_step}).  The UvN stable parameter
  values are shaded in gray, with the darker region corresponding to a
  single-domain system and the combined gray regions corresponding to
  a many-domain system.  The points represent numerical tests of
  gradient stability: the (blue) circles are parameter values
  that are stable for single-domain systems; these together with the
  (purple) squares are stable for multi-domain systems; and the
  $\times$ are parameter values that were found to be unstable.}
\label{fig:stability_diagram}
\end{center}
\end{figure}

Our primary results are the following: for the equation of motion,
Eq.~(\ref{eq:eq_of_motion}), there exists a class of first order,
semi-implicit steps
\begin{equation}
\begin{split}
h_{t+\Delta t} &= h_t + \Delta t\Bigl[-\nabla^4 h_t - \bm\nabla \cdot
\bigl\{(1-|\nabla h_t|^2) \bm\nabla h_t\bigr\}\Bigr] \\[1ex]
& +  b_1 \Delta t \nabla^2( h_{t+\Delta t} - h_t ) + b_2
 \Delta t \nabla^4( h_{t+\Delta t} - h_t)
\end{split}
\label{eq:discrete_step}
\end{equation}
that provides stable numerical integration for appropriate choice of
the parameters $b_1$ and $b_2$, as shown in
Fig.~\ref{fig:stability_diagram}.  The results of our UvN stability
analysis are presented as shaded regions while our numerical tests of
gradient stability are plotted as points.  Although UvN stability does
not ensure gradient stability, we find that it is very effective in
determining the gradient stable regions, both for single- and
many-domain systems.  The UvN stability conditions plotted here are
independent of lattice type or details of the numerical method (e.g.,
finite difference versus spectral methods).  The difference in
stability range for the single- and many-domain systems is revealed by
the UvN stability analysis, which shows that the most unstable Fourier
mode is that with its wavevector oriented with the local slope
$\bm\nabla h$.  In the many-domain system, each mode samples many
different slope directions, which acts to suppress the instability for
the parameter range $-1<b_1<-1/2$.

Our results are consistent with Xu and Tang, who proved gradient
stability for the parameters $b_1<-1$ and $b_2=1$ \cite{Xu06}.
Further work by one of us has led to a demonstration of gradient
stability for the entire dark gray shaded region of
Fig.~\ref{fig:stability_diagram} \cite{Vollmayr-Lee13}.  This proof
will be presented elsewhere, as it is considerably more general than
the model considered here, and does not distinguish the single- and
many-domain cases captured by the UvN stability analysis.

While our analysis is focused on the isotropic model,
Eq.~(\ref{eq:eq_of_motion}), our results can be generalized
straightforwardly to anisotropic growth, where only a discrete set of
slope orientations are preferred.  We demonstrate this explicitly for
a model with square symmetry, appropriate for growth on (100)
surface.

The remainder of the paper is as follows.  In
Sec.~\ref{sec:continuous_time} we review some of the properties of the
model to provide necessary background for subsequent sections.  In
Sec.~\ref{sec:UvN_stability} we present the UvN stability analysis,
both for single- and many-domain systems.  We describe the numerical
tests of gradient stability in Sec.~\ref{sec:numerical_tests}, as well
as providing the details of our finite-different implementation of
Eq.~(\ref{eq:discrete_step}).  In Sec.~\ref{sec:square_symmetry} we
extend our analysis to the anisotropic model with square symmetry.
This is followed by a summary in Sec.~\ref{sec:summary}.

\section{The Continuous Time Model}

\label{sec:continuous_time}

In this section we provide motivation for the model we are
considering and present some of its properties, showing in particular
the instability to pyramid formation and the energy decreasing dynamics
of the continuous time model.

The height field, $h(x,y,t)$, is defined in a co-moving frame so that
its average is zero, and obeys a continuity equation.  The current
$\bm J$ has an equilibrium surface diffusion contribution equal to the
gradient of the local curvature, $\bm
J_\text{SD}=\bm\nabla(\nabla^2h)$ \cite{Mullins59}, and a
non-equilibrium component $\bm J_\text{NE}$:
\begin{equation}
  \frac{\partial h}{\partial t} = -\bm\nabla\cdot\bm J = 
    -\nabla^4 h - \bm\nabla\cdot\bm J_\text{NE}.
  \label{eq:continuity_eq}
\end{equation}
A noise term is omitted as this is considered to be irrelevant for
coarsening \cite{Bray94}.  We consider the slope-selecting
nonequilibrium current
\begin{equation}
  \bm J_\text{NE} = (1 - |\nabla h|^2)\,\bm\nabla h, 
  \label{eq:j_ne}
\end{equation}
which gives $\bm J_\text{NE}\sim \bm\nabla h$ for small gradients, an
uphill current due to the Ehrlich-Schwoebel-Villain effect \cite{Villain91},
and $\bm J_\text{NE}=0$ for slopes of unit magnitude.  Inserting
Eq.~(\ref{eq:j_ne}) into the continuity equation
(\ref{eq:continuity_eq}) yields the equation of motion,
Eq.~(\ref{eq:eq_of_motion}).

Common variations on this model include slope-selecting currents that
vanish for only a discrete set of $\bm\nabla h$ directions, reflecting
the underlying crystalline structure, and models without slope
selection.  The physical basis and experimental evidence for these
various models is described in
\cite{Johnson94,Siegert94,Rost97,Ortiz99,Moldovan00} and references
therein.  Material parameters have been absorbed into rescaling of
lateral space dimensions, height, and time.

The equation of motion, Eq.~(\ref{eq:eq_of_motion}),
can be written as a gradient flow
\begin{equation}
\frac{\partial h}{\partial t} = -\frac{\delta F}{\delta h}
\end{equation}
 for the free energy functional
\begin{equation}
  F[h] = \int d^2x \left\{{\textstyle\frac{1}{2}}(\nabla h)^2 + 
  {\textstyle\frac{1}{4}} \left(1- |\nabla h|^2\right)^2\right\}.
  \label{eq:free_energy}
\end{equation}
Gradient flow results in a monotonically decreasing free
energy,
\begin{equation}
\frac{d}{dt} F = \int d^2x \left(\frac{\delta F}{\delta h}\right) 
\frac{\partial h}{\partial t}
 = -\int d^2x \left(\frac{\partial h}{\partial t}\right)^2 \leq 0.
\label{eq:decreasing_free_energy}
\end{equation}
As first noted by Eyre \cite{Eyre98b}, the essential stability
criterion for discrete time steps is to preserve the energy decreasing
property of the continuous-time equation.  This is known as the
gradient stability condition.

Next we review the the linear stability of the continuous time
equation, which will be useful context for the von Neumann stability
analysis in Sec.~\ref{sec:UvN_stability}.  Consider a height field
\begin{equation}
h(x,y,t) = C x + \eta(x,y,t),
\label{eq:fluctuations}
\end{equation}
which consists of small deviations $\eta$ from a uniform slope.
Inserting this into Eq.~(\ref{eq:eq_of_motion}), linearizing in
$\eta$, and Fourier transforming to $\tilde\eta({\bf k},t)\equiv \int d^2x\,
\exp(i{\bf k}\cdot{\bf x}) \eta(x,y,t)$ gives
\begin{equation}
  \frac{\partial\tilde\eta({\bf k},t)}{\partial t} = (k^2 - k^4
  - C^2 k^2 - 2 C^2 k_x^2 )\, \tilde\eta({\bf k},t).
\end{equation}
For an interface that is initially flat we set $C=0$ and obtain
the growth rate for small fluctuations in the initial conditions:
\begin{equation}
  \frac{\partial\tilde\eta({\bf k},t)}{\partial t} = 
   k^2 (1-k^2)\,\tilde\eta({\bf k},t).
\end{equation}
Long wavelength modes with $k<1$ are unstable and grow, which is
exactly the instability that leads to pyramid formation.  In the
context of the Cahn-Hilliard equation this is the spinodal
instability \cite{Bray94}.  Note that the exponential growth of the
mode is nevertheless accompanied by a decreasing total free energy, as
required by Eq.~(\ref{eq:decreasing_free_energy}).

For an equilibrium interface we set the slope $C=1$ to
obtain
\begin{equation}
  \frac{\partial\tilde\eta({\bf k},t)}{\partial t} = 
   -(k^4 + k_x^2)\,\tilde\eta({\bf k},t).
\end{equation}
The negative right hand side indicates that height fluctuations about
the equilibrium slope decay, and the uniform slope profile is stable.

\section{Unconditional von Neumann Stability Analysis}

\label{sec:UvN_stability}

The goal in constructing a discrete time method is to be faithful to
the physical behavior of the continuous time equation.  In our case,
this means our discrete step should be gradient stable, to preserve
the energy-decreasing property of the continuous equation.  However,
in this section we analyze instead von Neumann (vN) stability,
i.e.\ the linear stability of the discrete step,
Eq.~(\ref{eq:discrete_step}).  This analysis has certain advantages.
It is relatively straightforward and, as shown in
Fig.~\ref{fig:stability_diagram} and in Ref.~\cite{Vollmayr-Lee03}, it
successfully predicts the parameter range for gradient stability, as
judged by numerical tests.  Also, the method provides insight into the
dynamics of the Fourier modes, which in the present case proves useful
in clarifying the distinction between the single- and many-domain
systems.

We first present vN stability analysis on the Euler step, which
results in conditional stability, i.e.,\ a lattice-dependent upper
bound on $\Delta t$.  Then we consider our parametrized semi-implicit
step and perform unconditional vN stability analysis; that is, we
seek parameter values which yield vN stable steps for any size
$\Delta t$.  Note that we will only impose vN stability on the 
equilibrium, sloped interface and not on the flat interface,
where the linear instability is part of the physical behavior
of the continuum equation.

In addition to the time discretization, the spatial derivatives in our
equation of motion must be treated by finite-difference or spectral
methods.  Without specifying the details of the scheme, we denote the
Fourier transform of the two-dimensional numerical laplacian as
$\lambda({\bf k})$.  In the continuum limit, $\lambda({\bf k})\to
-k^2$.  For spatially discretized systems, $0 \geq \lambda({\bf k})
\geq \lambda_\text{min}$, where the value of the lower bound
$\lambda_\text{min}\sim -1/\Delta x^2$ depends on the details of
the discretized laplacian.  Our stability conditions will rely
only on the universal upper bound of zero.

We will use $\lambda({k_x})$ to represent the Fourier transform of the
numerical derivative second derivative $\partial^2/\partial x^2$.

\subsection{Euler Step}

Our discrete time step, Eq.~(\ref{eq:discrete_step}), reduces to an
Euler step in the case $b_1=b_2=0$.  We plug in $h=x+\eta$ (i.e.,\ slope
$C=1$),
linearize in $\eta$, and Fourier transform to obtain
\begin{equation}
  \tilde\eta_{t+\Delta t} = 
 \Bigl[ 1 + \Delta t \Bigl\{-\lambda({\bf k})^2 + 2  \lambda(k_x)
  \Bigr\} \Bigr] \tilde\eta_t.
\label{eq:euler_sloped}
\end{equation}
The vN stability condition is that the square bracket term has
magnitude less than unity, to ensure fluctuations die away.  The
negative curly bracket term in Eq.~(\ref{eq:euler_sloped}) has no
lower bound in the continuum limit $\Delta x\to 0$, and thus the Euler
step would be vN unstable for any size $\Delta t$.  The situation is
improved by the numerical derivative, which places a lower bound on
the curly bracket terms, leading to vN stability for $\Delta t
\lesssim |\lambda_\text{min}|^{-2}\sim \Delta x^4$.  The analysis is
essentially identical to what happens in the Cahn-Hilliard equation
\cite{Rogers88}.  The Euler step provides an example of a
lattice-dependent stability condition (relying on the lower bound of
$\lambda({\bf k})$ rather than the upper bound of zero) and it results
in a fixed bound on the time step, regardless of the natural time
scale of the dynamics.

\subsection{UvN Stability for a Single Domain}

We return to our parametrized discrete step,
Eq.~(\ref{eq:discrete_step}), but now we leave $b_1$ and $b_2$
unspecified.  We seek to find ranges for the parameters which will
lift any restrictions on $\Delta t$, i.e.,\ unconditional
stability.  We substitute Eq.~(\ref{eq:fluctuations}) with slope 
$C=1$ into Eq.~(\ref{eq:discrete_step}), linearize,
and Fourier transform.  The resulting step can be written as
\begin{equation}
[1 + \Delta t {\cal L}({\bf k})] \, \tilde\eta_{t+\Delta t} = 
[1 + \Delta t {\cal R}({\bf k})] \, \tilde\eta_{t} 
\label{eq:UvN_LR_eq}
\end{equation}
with
\begin{equation}
{\cal L}({\bf k}) = b_1\lambda({\bf k}) + b_2\lambda({\bf k})^2
\end{equation}
and 
\begin{equation}
{\cal R}({\bf k}) = 2\lambda(k_x) +b_1\lambda({\bf k})
 + (b_2-1) \lambda({\bf k})^2.
\end{equation}
Before imposing the UvN stability, we note that it is necessary to
have ${\cal L}({\bf k}) \geq 0$ so that the square bracket on the left
of Eq.~(\ref{eq:UvN_LR_eq}) is non-vanishing for all $\Delta t$ and
${\bf k}$.  This gives the requirement that $b_1\leq 0$ and $b_2\geq
0$.  

Next, the UvN stability condition, $|\tilde\eta_{t+\Delta t}| <
|\tilde\eta_t|$ for all $\Delta t$ and ${\bf k}$, will be
satisfied if ${\cal L}({\bf k}) > |{\cal R}({\bf k})|$.  In the case
that ${\cal R}({\bf k})$ is positive, this gives the condition
\begin{equation}
0 < {\cal L}({\bf k}) - {\cal R}({\bf k}) = 
 -2\lambda(k_x)+\lambda({\bf k})^2,
\end{equation}
which is intrinsically satisfied due to the non-positivity of
$\lambda(k_x)$.  While here and below the ${\bf k}=0$ mode saturates
the bound, we can safely ignore it since it is static.

The crucial condition, then, comes from imposing ${\cal L}({\bf k})
 > -{\cal R}({\bf k})$, which becomes
\begin{equation}
 \lambda(k_x) + b_1\lambda({\bf k}) + \left(b_2 -\frac{1}{2}\right)
\lambda({\bf k})^2 > 0.
\label{eq:crucial_condition}
\end{equation}
The last term is positive for $b_2>1/2$.  Next, noting that
$\lambda(k_x) \geq \lambda({\bf k})$, we have a lower bound on the
remaining two terms:
\begin{equation}
\lambda(k_x) + b_1\lambda({\bf k}) \geq (1+b_1)\lambda({\bf k}).
\end{equation}
This will be positive provided that $b_1 < -1$. 
 Thus, our conditions
for UvN stability of a single-domain system are
\begin{equation}
b_1<-1, \quad b_2>1/2,
\end{equation}
which is plotted as the dark gray region of
Fig.~\ref{fig:stability_diagram}. 

Note that for $b_1$ slightly above $-1$, in the unstable region, it is
Fourier modes with $\lambda(k_x)\approx\lambda({\bf k})$ that first
violate Eq.~(\ref{eq:crucial_condition}).  This corresponds to wavevectors
${\bf k}$ that are nearly oriented along the $x$-axis, i.e.\ the
gradient direction of the equilibrium interface.

\subsection{UvN Stability for a Many-Domain System}

In a many-domain system, which is the relevant case for coarsening
studies, we are not free to choose the coordinate axes to align the
$x$ axis with the interface gradient, since there are many facets with
different gradient directions.  To analyze this case, we first
linearize about a single domain but with an arbitrary normal
direction, parametrized by the polar coordinate $\theta$
\begin{equation}
h(x,y,t) =  \cos(\theta) x + \sin(\theta) y + \eta(x,y,t).
\end{equation}
This follows through just as before, with the important stability
condition Eq.~(\ref{eq:crucial_condition}) becoming
\begin{equation}
\cos^2\theta \lambda(k_x) + \sin^2\theta\lambda(k_y) + b_1\lambda({\bf k})
 + \left(b_2-\frac{1}{2}\right) \lambda({\bf k})^2 > 0.
\label{eq:many_domains}
\end{equation}
Now, if many domains are present in the system with essentially random
orientations, then for any particular Fourier mode the above equation
will be averaged over $\theta$, giving $\langle\cos^2\theta \rangle =
\langle\sin^2\theta\rangle = 1/2$.  Using
\begin{equation}
\lambda(k_x) + \lambda(k_y) \approx \lambda({\bf k})
\label{eq:laplacian_relation}
\end{equation}
reduces Eq.~(\ref{eq:many_domains}) to
\begin{equation}
\left(b_1 + \frac{1}{2} \right) \lambda({\bf k}) +
\left(b_2-\frac{1}{2}\right) \lambda({\bf k})^2 > 0.
\end{equation}
Thus, our UvN stability condition for many-domain systems is
\begin{equation}
 b_1 < -1/2, \quad b_2> 1/2,
\end{equation}
which is depicted as the combined shaded regions of
Fig.~\ref{fig:stability_diagram}.  The averaging over multiple
orientations provides a greater parameter range of stability than the
single-domain case.

Note that in general Eq.~(\ref{eq:laplacian_relation}) is only an
approximate relationship.  It is a strict equality in the $\Delta x\to
0$ continuum limit, and also in the common five-point stencil for the
numerical laplacian on a square lattice, but for other choices of
numerical derivatives it need not be exact.

\section{Numerical Tests of Gradient Stability}

\label{sec:numerical_tests}

Since the field equation of motion is nonlinear, von Neumann stability
analysis is not sufficient to prove gradient stability.  For that
reason, we have conducted extensive numerical tests for gradient
stability for a range of $b_1$ and $b_2$ parameter values.  We present
the details of the numerical derivative implementation in an appendix,
but we note here two important general features such an implementation
should have.  First, the local conservation law should be constructed
to hold exactly, not just to some order in $\Delta x$, and second, the
energy-decay property of the continuous time equation should be
maintained when spatially discretizing.  That is, the particular
scheme of calculating the spatially discrete analog of the free energy
$F[h]$ in Eq.~(\ref{eq:free_energy}) and the equation of motion should
be consistent, so that
\begin{equation}
\frac{d}{dt} h_{ij} = -\frac{\partial}{\partial h_{ij}}
  \left(\frac{F}{\Delta x^2}\right)
\end{equation}
is an exact relation, not just approximate to some order in 
$\Delta x$.

For each $b_1$ and $b_2$ value represented as a data point in
Fig.~\ref{fig:stability_diagram} we performed the following tests.  We
evolved a $512\times 512$ sized lattices with lattice constant $\Delta
x=1$ out to a final time $t_\text{max}$.  These systems were evolved
using three different methods: an Euler step with $\Delta t = 0.03$
out to a $t_\text{max}=10^4$, a semi-implicit step with $b_1=2.5$ and
$b_2=1$ and growing time step $\Delta t = 0.03 t^{2/3}$ out to time
$t_\text{max}=10^6$, and the same semi-implicit parameters with a
fixed time step $\Delta t=100$ out to time $t_\text{max}=10^6$.  For
each of these cases we analyzed multiple runs and varied between
random initial conditions and sinusoidal initial conditions with long
and short wavelengths.

At regular intervals during the evolution we tested a single step
calculated via Eq.~(\ref{eq:discrete_step}) with sizes varied between
$1 \leq \Delta \leq 10^{10}$.  This step was used only for energy
stability testing and did not contribute to the subsequent time
evolution.  Any time that the free energy was found to increase, that
particular set of parameter values was identified as unstable.

For the many-domain system, we used periodic boundary conditions and
an initially flat interface (plus the random or sinusoidal
fluctuations).  For the single-domain system, we first re-write the
field equation of motion, Eq.~(\ref{eq:eq_of_motion}) in terms of
deviations from the uniform slope, giving
\begin{equation}
\begin{split}
\frac{\partial\eta}{\partial t} = 
&-\nabla^4\eta +2 \partial_x^2\eta
 + 2 \partial_x |\nabla\eta|^2
 + 2 (\partial_x\eta) \nabla^2\eta\\
 & + \bm\nabla\cdot(|\nabla\eta|^2\bm\nabla\eta),
\end{split}
\end{equation}
where $\partial_x = \partial/\partial x$, and then constructed the
analogous numerical implementation of this equation.  This approach
was necessary to eliminate sensitivity to truncation error.  We
imposed periodic boundary conditions on $\eta$, which corresponds to
shifted periodic boundary condition on $h$.

In Fig.~\ref{fig:stability_diagram} we show the results of this
testing both for the single- and many-domain systems.  The (blue)
circles represent parameter values that were found to be stable for
the single-domain system, that is, under all our testing, there were
no single incidents of energy increase.  The (purple) squares are
parameters values that were found to be unstable in the single-domain
system, but stable for the many-domain case.  The remaining $\times$
are parameter values found to be unstable for both single- and
many-domain systems.
We find a striking degree of agreement between the predictions of
UvN stability analysis and the numerical tests for unconditional
gradient stability.  This is one of our main results.  

There is a small region for $b_2<1/2$ where numerical tests find
gradient stability.  This can be understood from
Eq.~(\ref{eq:crucial_condition}) as a lattice-dependent stability
arising from the laplacian lower bound $\lambda_\text{min}$.  We have
emphasized instead the lattice-independent stability boundaries, as
these are more widely applicable.

\begin{figure}
\begin{center}
\includegraphics[width=3in]{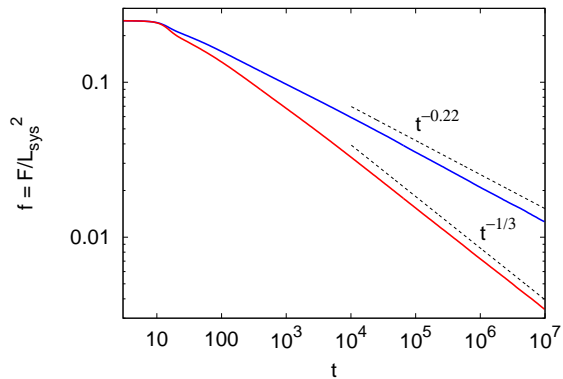}
\caption{(color online) The free energy density $f = F[h]/L_\text{sys}^2$ as
a function of time, where the time evolution utilized a growing
time step, $\Delta t\sim t^{2/3}$.  Simulation details are in the
text.  The lower (red) curve is the isotropic model, while the
upper (blue) curve is for the anisotropic model with square
symmetry presented in Sec.~\ref{sec:square_symmetry}.}
 \label{fig:free_energy}
\end{center}
\end{figure}

To illustrate the utility of these methods, we have simulated the
coarsening that results from an initially flat interface, using a
the stable step parameters $b_1=-1.5$ and $b_2=1$ and a
growing step size $\Delta t = \max(0.1, 0.01 t^{2/3})$.  We performed 20
independent runs on a $2048\times 2048$ lattice with $\Delta x=1$, out
to time $t_\text{max}=10^7$.

Fig.~\ref{fig:free_energy} shows the decay of the free energy with
time.  Once equilibrated domains form, the free energy density $F$ is
proportional to the amount of edge in the system, which is inversely
proportional to the characteristic size of the domains.  Thus the free
energy should decay as $F\sim 1/L(t) \sim t^{-1/3}$.  Our growing time step
integration reproduces this result.

\begin{figure}
\begin{center}
\includegraphics[width=4.2in]{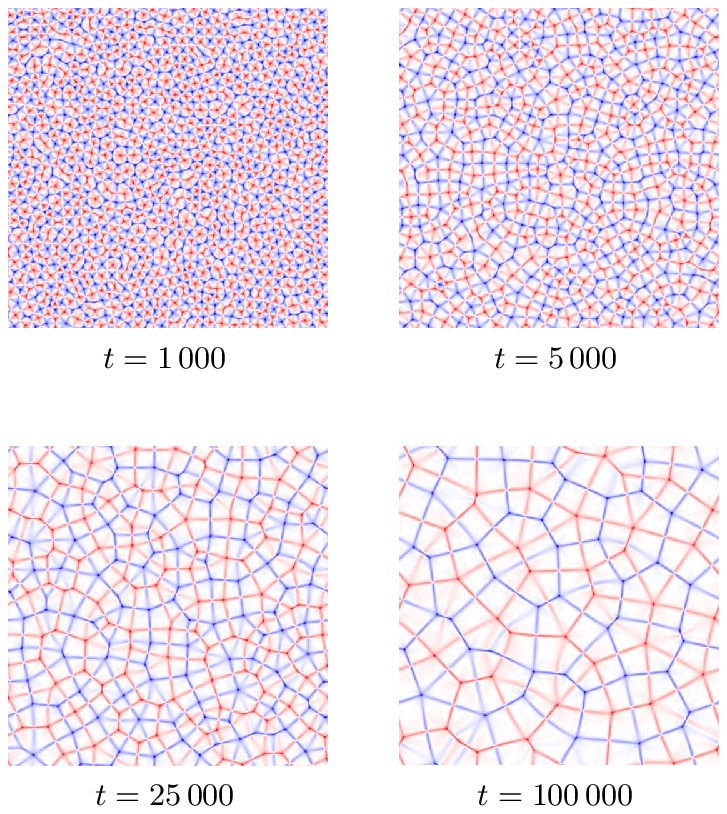}
\caption{(color online) Plotted is the laplacian of $h(x,y,t)$, for a
  system evolved with a growing time step $\Delta t\sim t^{2/3}$.
  Simulation details are provided in the text.  Positive values
  (troughs) are red, negative values (peaks) are blue, and the white
  regions are domains of uniform slope with zero laplacian. }
\label{fig:domains}
\end{center}
\end{figure}

Shown in Fig.~\ref{fig:domains} are snapshots of domain configurations
for various times from a single run on a $512\times 512$ lattice, with
all other parameters as given above.

\section{Model with Square Symmetry}

\label{sec:square_symmetry}

While the isotropic growth model, Eq.~(\ref{eq:eq_of_motion}),
provides a useful starting point for analyzing surface growth
coarsening, experimental systems typically select for only a discrete
set of slope orientations.  For example, homoepitaxial growth on a
Cu(100) surface exhibits a square symmetry with four equilibrium slope
orientations \cite{Zuo97}. 
This symmetry can be easily added to the
phase-field model by adding a term to the free energy functional
\begin{equation}
  F_\text{sq}[h] = F_\text{iso}[h] + \int d^2x \,
  (\partial_x h)^2 (\partial_y h)^2
  \label{eq:free_energy4}
\end{equation}
where $F_\text{iso}[h]$ is the free energy of
Eq.~(\ref{eq:free_energy}), and $\partial_x = \partial/\partial x$.
The additional term is non-negative and vanishes for slopes oriented
with the cartesian axes.  We choose a prefactor of unity for this term
since this results in an isotropic potential to quadratic order about
any of the four equilibrium points.

Taking $\partial h/\partial t = -\delta F_\text{sq}/\delta h$ then gives
the equation of motion
\begin{align}
\frac{\partial h}{\partial t} = -\nabla^4 h &-
\partial_x \left\{ \left[ 1 - |\nabla h|^2 - 2(\partial_y h)^2
 \right] \partial_x h\right\} \nonumber\\
&- \partial_y \left\{ \left[ 1 - |\nabla h|^2 - 2(\partial_x h)^2
\right] \partial_y h\right\}.
\end{align}
We parametrize our first order accurate time step as before, with
\begin{equation}
\begin{split}
h_{t+\Delta t} = h_t + \Delta t \left(\frac{\partial h}{\partial t}\right)_t
 &+  b_1 \Delta t \nabla^2( h_{t+\Delta t} - h_t ) \\
 &+ b_2 \Delta t \nabla^4( h_{t+\Delta t} - h_t).
\end{split}
\end{equation}
UvN stability analysis about an equilibrium slope, $h=x+\eta$,
takes the same form Eq.~(\ref{eq:UvN_LR_eq}), with ${\cal L}({\bf k})$
unchanged and
\begin{equation}
{\cal R}({\bf k}) = (2+b_1)\lambda({\bf k}) + (b_2-1)\lambda({\bf k})^2.
\end{equation}
The crucial condition ${\cal L}+{\cal R}>0$ then results in the 
stability region
\begin{equation}
b_1 < -1, \qquad b_2 > 1/2,
\end{equation}
with no distinction between single and multiple domain systems.
We conducted numerical tests of gradient stability following the
same protocol shown in Sec.~\ref{sec:numerical_tests} and again
find good agreement, as shown in Fig.~\ref{fig:stability_diagram4}.
The details of our numerical spatial derivatives are provided in the
appendix.

\begin{figure}
\begin{center}
\includegraphics[width=3.25in]{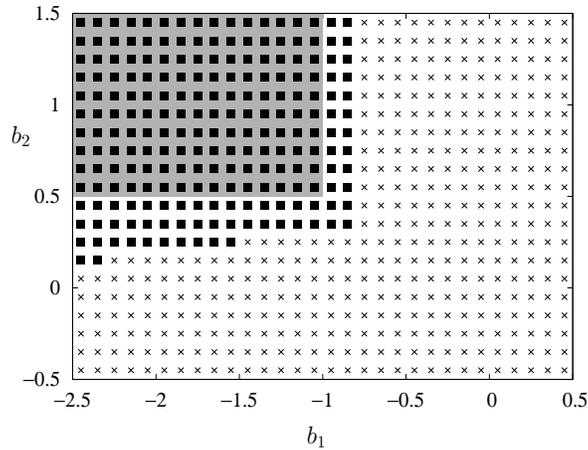}
\caption{Stability diagram for the square symmetry model of
  Sec.~\ref{sec:square_symmetry}.  Squares represent $(b_1, b_2)$
  parameter values which were gradient stable in our numerical tests,
  whereas the $\times$ were found to be unstable.  The shaded region
  represents UvN stable parameter values..}
\label{fig:stability_diagram4}
\end{center}
\end{figure}

\begin{figure}
\begin{center}
\includegraphics[width=4.2in]{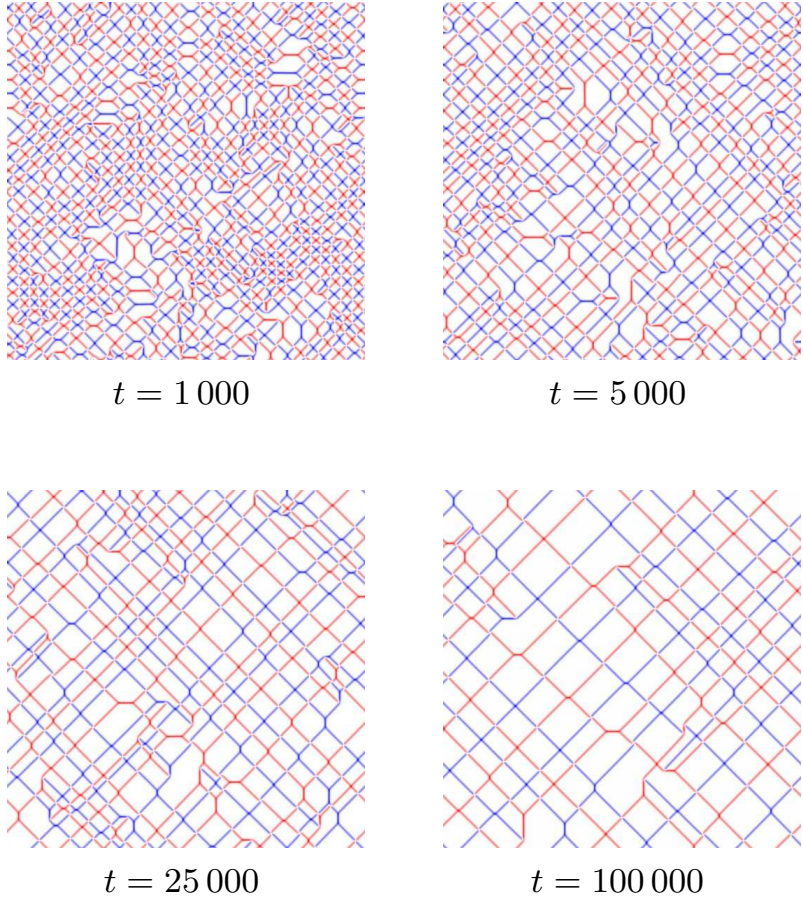}
\caption{(color online) Plotted is the laplacian of $h(x,y,t)$ as
  described in Fig.~\ref{fig:domains}, but here for the square symmetry
  model.}
\label{fig:domains4}
\end{center}
\end{figure}

Experiments \cite{Zuo97} and simulations \cite{Siegert98,Moldovan00}
find $L\sim t^{1/4}$ growth for crystal growth with square symmetry
(although variants of this square symmetry model can result in
$t^{1/3}$ growth \cite{Levandovsky04}).  We measured the length scale
via the free energy density following the same procedure as described
in Sec.~\ref{sec:numerical_tests}, and the results are presented in
Fig.~\ref{fig:free_energy}.  For the time range simulated, we observe
slightly slower than $t^{1/4}$ growth, with a exponent around $0.22$.
Finally, we show in Fig.~\ref{fig:domains4} snapshots of typical
domain configurations from a single run on a $512\times 512$ lattice.

As this section demonstrates, it is straightforward to generalize
the analysis of the isotropic model to the case with a discrete
set of preferred slope orientations.  In particular, the analysis
for models with six-fold symmetry \cite{Moldovan00} and three-fold
symmetry \cite{Watson06} should follow analogously.

\section{Summary}

\label{sec:summary}

We have parametrized a first order accurate discrete time step,
Eq.~(\ref{eq:discrete_step}) for MBE growth with slope selection that,
unlike the Euler step, is gradient stable for appropriate choices of
the parameters $b_1$ and $b_2$.  We determined the stability range for
these parameters via unconditional von Neumann stability analysis, and
then tested these predictions with numerical tests for gradient
stability, as shown in Fig.~\ref{fig:stability_diagram}.  We find
that the UvN stability analysis serves as an accurate proxy for
unconditional gradient stability, similar to the behavior of the 
Cahn-Hilliard equation \cite{Vollmayr-Lee03}.

Our stability analysis contained an implicit assumption that the
interface slopes do not exceed unit magnitude, which we justify by
noting that the dynamics naturally select for this slope.  This came
into our UvN analysis by our choice to linearize about a unit slope
domain.  We note that the numerical tests for gradient stability
contained no such assumption, so the agreement between the two
approaches confirms validity of the unit slope assumption.

The increase in efficiency due to a gradient stable method is
substantial.  For the simulations presented in
Fig.~\ref{fig:free_energy}, computation by Euler step, for which the
largest stable step size is $\Delta t=0.03$, would require $3.3\times
10^8$ time steps.  In contrast, using a stable method with step size
$\Delta t= \max(\Delta t_0,At^{2/3})$ the number of time steps
required to reach some $t_\text{max}$ is given by
$3t_\text{max}^{1/3}/A$, which for our simulations is $6.5\times 10^4$
steps.  Each stable step involves an overhead factor of 2.4 due to the
addition of the FFT, but the net result is an overall increase of
efficiency by a factor of 2100 for the data we present!  Note
that this factor will increase as computational resources allow for
larger systems to be evolved to later times.

The method of parametrizing linear semi-implicit steps, performing
unconditional von Neumann stability analysis, and then testing
the predictions numerically for gradient stability has yielded
efficient stable methods for the Cahn-Hilliard and Allen-Cahn
equations \cite{Vollmayr-Lee03} and now for a class of MBE 
crystal growth models.  We anticipate that this procedure will
prove useful to many other phase field models.

\section*{Acknowledgments}

G.M.S. was supported by NSF REU
Grant~PHY-1156964. B.P.V.-L. acknowledges financial support from the
Max Planck Institute for Dynamics and Self-Organization and the
hospitality of the University of G\"ottingen, where this work was
completed.




\appendix

\section{Finite Difference Scheme}

\label{appendix}

Here we present details of the spatial discretization scheme we used
in our numerical tests.  We present these in a discrete-space, continuous
time  picture, as our goal is to
ensure that the conservative dynamics and the gradient flow are
exact, i.e. preserved to all orders in $\Delta x$.  The essential
condition for gradient flow is that the equation of motion must be
connected to a particular choice for the free energy functional such
that
\begin{equation}
\frac{\partial h_{i,j}}{\partial t} = -\frac{\partial}{\partial h_{i,j}}
\left(\frac{F}{\Delta x^2}\right).
\label{eq:discrete_gradient_flow}
\end{equation}
Local conservation is imposed by ensuring that the equation of motion
has the form
\begin{equation}
\begin{split}
\frac{dh_{i,j}}{dt} = -\frac{1}{\Delta x}&\Bigl[ 
  \{J_x\}_{i+1/2,j}- \{J_x\}_{i-1/2,j} \\
& - \{J_y\}_{i,j+1/2}- \{J_y\}_{i,j-1/2}\Bigr]
\end{split}
\label{eq:discrete_continuity_eq}
\end{equation}
so that the same $\{J_x\}_{i+1/2,j}$ flows into $h_{i+1,j}$ and out of
$h_{i,j}$, and the same $\{J_y\}_{i,j+1/2}$ flows into $h_{i,j+1}$ and
out of $h_{i,j}$.

Our implementation uses an on-site finite-difference expression for
$\nabla^2 h$, for which we take the standard five-point stencil,
\begin{equation}
\{\nabla^2h\}_{i,j} = \frac{1}{\Delta x^2}[h_{i+1,j} +
  h_{i-1,j} + h_{i,j+1}+ h_{i,j-1}-4h_{i,j}],
\end{equation}
and the cell-centered expression for $|\nabla h|^2$,
\begin{equation}
\begin{split}
\{|\nabla h|^2&\}_{i+1/2,j+1/2} = \\ \frac{1}{2\Delta x^2}&
\Bigl[(h_{i+1,j}-h_{i,j})^2 + (h_{i+1,j+1}-h_{i,j+1})^2  \\ &+ 
(h_{i,j+1}-h_{i,j})^2 + (h_{i+1,j+1}-h_{i+1,j})^2\Bigr].
\end{split}
\end{equation}
With these choices it is straightforward to show that
\begin{equation}
\frac{\partial}{\partial h_{k,l}} \sum_{i,j} \{ |\nabla h|^2 \}_{i+1/2,j+1/2}
 = - 2\{\nabla^2 h\}_{k,l}.
\label{eq:derivative_relation}
\end{equation}

Our equation of motion is given by Eq.~(\ref{eq:discrete_gradient_flow})
with the choice
\begin{equation}
\frac{F}{\Delta x^2} = \sum_{i,j} \left[ \frac{1}{2}\{\nabla^2 h\}_{i,j}^2
 + \frac{1}{4}\Bigl(1-\{|\nabla h|^2\}_{i+1/2,j+1/2}\Bigr)^2\right].
\label{eq:discrete_F}
\end{equation}
By making use of Eq.~(\ref{eq:derivative_relation}), the equation
of motion can be shown to satisfy the discrete continuity equation 
(\ref{eq:discrete_continuity_eq}) with current
\begin{equation}
\{J_x\}_{i+1/2,j} = \{J^{SD}_x\}_{i+1/2,j} + \{J^{NE}_x\}_{i+1/2,j} 
\end{equation}
where the surface diffusion current is
\begin{equation}
  \{J^{SD}_x\}_{i+1/2,j} = \frac{\{\nabla^2 h\}_{i+1,j}-\{\nabla^2 h\}_{i,j}}{\Delta x},
\end{equation}
and the nonequilibrium current is
\begin{equation}
\begin{split}
\{&J^{NE}_x\}_{i+1/2,j} = \frac{h_{i+1,j}-h_{i,j}}{\Delta x}\\
 &\times\left[1 - \frac{1}{2}\Bigl(\{|\nabla h|^2\}_{i+1/2,j+1/2}
 + \{|\nabla h|^2\}_{i+1/2,j-1/2}\Bigr)\right],
\end{split}
\end{equation}
and analogous expressions for $\{J_y\}_{i,j+1/2}$.
The discrete from of the free energy, Eq.~(\ref{eq:discrete_F}), was
used the numerical tests for gradient stability.

For the square symmetry model, we need additionally the 
cell-centered derivatives
\begin{equation}
\begin{split}
\{(\partial_x h)^2&\}_{i+1/2,j+1/2} = \\ &\frac{1}{2\Delta x^2}
\Bigl[(h_{i+1,j}-h_{i,j})^2 + (h_{i+1,j+1}-h_{i,j+1})^2\Bigr] \\[2ex]
\{(\partial_y h)^2&\}_{i+1/2,j+1/2} = \\ &\frac{1}{2\Delta x^2}
\Bigl[(h_{i,j+1}-h_{i,j})^2 + (h_{i+1,j+1}-h_{i+1,j})^2\Bigr].
\end{split}
\end{equation}
The free energy is given by Eq.~(\ref{eq:discrete_F}) with the
additional term
\begin{equation}
\frac{F_\text{sq}}{\Delta x^2} = 
\frac{F}{\Delta x^2} + \sum_{i,j} \{(\partial_x h)^2\}_{i+1/2,j+1/2}
\{(\partial_y h)^2\}_{i+1/2,j+1/2}
\end{equation}
which corresponds to the nonequilibrium currents
\begin{equation}
\begin{split}
\{J^{NE,sq}_x&\}_{i+1/2,j} = \{J^{NE}_x\}_{i+1/2,j} -
 \frac{h_{i+1,j}-h_{i,j}}{\Delta x}\\
 &\times \Bigl(\{(\partial_y h)^2\}_{i+1/2,j+1/2}
 + \{(\partial_y h)^2\}_{i+1/2,j-1/2}\Bigr),
\end{split}
\end{equation}
and
\begin{equation}
\begin{split}
\{J^{NE,sq}_y&\}_{i,j+1/2} = \{J^{NE}_y\}_{i,j+1/2} -
 \frac{h_{i,j+1}-h_{i,j}}{\Delta x}\\
 &\times \Bigl(\{(\partial_x h)^2\}_{i+1/2,j+1/2}
 + \{(\partial_x h)^2\}_{i-1/2,j+1/2}\Bigr).
\end{split}
\end{equation}


\bibliographystyle{elsarticle-num}
\bibliography{bvl.bib}







\end{document}